\documentclass[pra,showpacs,amsmath,amssymb,twocolumn,superscriptaddress]{revtex4}
\usepackage{bm,multirow}
\usepackage[usenames]{color}
\newcommand \beq{\begin{eqnarray}}
\newcommand \eeq{\end{eqnarray}}
\def\simge{\mathrel{%
       \rlap{\raise 0.511ex \hbox{$>$}}{\lower 0.511ex \hbox{$\sim$}}}}
\def\simle{\mathrel{
       \rlap{\raise 0.511ex \hbox{$<$}}{\lower 0.511ex \hbox{$\sim$}}}}
\usepackage{graphics}
\usepackage[dvips]{graphicx}
\usepackage{graphicx}
\newcommand{\qq}{\mathbf{q}}
\newcommand{\pp}{\mathbf{p}}

\newcommand{\vv}{\mathbf{v}}

\newcommand{\be}{\begin{equation}}\newcommand{\ee}{\end{equation}}
\newcommand{\bea}{\begin{eqnarray}}
\newcommand{\eea}{\end{eqnarray}}
\newcommand{\ba}{\begin{align}}
\newcommand{\ea}{\end{align}}

\newcommand{\rmi}{{\rm i}}

\begin{document}
\title{Induced interactions in dilute atomic gases and liquid helium mixtures}
\author{Zhenhua Yu}
\affiliation{The Niels Bohr International Academy, The Niels Bohr Institute, Blegdamsvej 17, DK-2100 Copenhagen \O, Denmark}
\affiliation{Department of Physics, The Ohio State University, 191 West Woodruff Avenue,
Columbus, Ohio 43210-1117}
\affiliation{Institute for Advanced Study, Tsinghua University, Beijing 100084, China}
\author{C.\ J.\ Pethick}
\affiliation{The Niels Bohr International Academy, The Niels Bohr Institute, Blegdamsvej 17, DK-2100 Copenhagen \O, Denmark}
\affiliation{NORDITA,   Royal Institute of Technology and Stockholm University,   Roslagstullsbacken 23, SE-10691 Stockholm, Sweden}

\date{\today}

\begin{abstract}
In dilute mixtures of two atomic gases, interactions between  two minority atoms acquire a contribution due to interaction with the majority component.  Using thermodynamic arguments, we derive expressions for this induced interaction for both fermions and bosons for arbitrary strength of the interaction between the two components.    Implications of the work for the theory of dilute solutions of $^3$He in liquid $^4$He are discussed.


\end{abstract}

\maketitle

\section{Introduction}

Induced interactions are responsible for a variety of phenomena in condensed matter physics, ranging from the superconductivity of metals, to the stability of the A phase of superfluid liquid $^3$He, to the effective interactions between $^3$He atoms in liquid $^4$He \cite{BBP}.  Recently, following the work of Mora and Chevy \cite{mora}, it was shown that such processes are important in dilute mixtures of atomic gases \cite{YZP}.  In this paper, we expand on Ref. \cite{YZP}, which considered two fermonic species, and extend the results to boson-fermion and boson-boson mixtures. In addition, we draw a number of conclusions relevant to dilute solutions of $^3$He in liquid $^4$He.

\section{Fermion--fermion mixtures}
\label{Sec:FF}

We begin by giving a compact derivation of the basic result that the Landau quasiparticle interaction for two minority  atoms (denoted by 2) in a Fermi gas consisting of a majority species 1, has a momentum independent contribution of the form
\be
f= \nu^2 \frac{\partial \mu_1}{\partial n_1}.
\label{f1}
\ee
Here
\be
\nu=\left.\frac{\partial n_1}{\partial n_2}\right|_{\mu_1}
\label{nu}
\ee
is the number of majority atoms that must be added per minority atom in order to keep the chemical potential of the majority species fixed and $\partial n_1/\partial \mu_1={\cal N}_1(0)=m_1p_{F1}/2\pi^2$ is the density of single-particle states at the Fermi surface for species 1. (We use units in which $\hbar$ is equal to unity.) The density of species $i$ is denoted by $n_i$, the mass of an atom by $m_i$, and the Fermi momentum by $p_{Fi}$.  Physically, the quantity $\nu$ is the number of majority atoms in the dressing cloud of a minority atom.

To derive the result (\ref{f1}), it is simplest to work in terms of an effective low energy theory, in which high-lying states are eliminated, and only low-lying ones are retained.   The effective low-energy Hamiltonian is
\begin{align}
H=&\sum_{\pp_1}  \frac{p_1^2}{2 m_1} a^\dagger_{\pp_1}   a_{\pp_1} +\epsilon_2N_2 +\sum_{\pp_2}  \frac{p_2^2}{2 m_2^*} b^\dagger_{\pp_2}   b_{\pp_2}\nonumber\\
&
+\frac{g}{V} {\sum_{\pp_1\pp_2\qq}}'    a^\dagger_{\pp_1+\qq} b^\dagger_{\pp_2-\qq} b_{\pp_2} a_{\pp_1},
\label{ham}
\end{align}
where $V$ is the volume of the system, $N_1$ is the total number of 1-atoms, $\epsilon_2$ is the energy to add a single 2-atom to the 1-atoms,  $m_2^*$ is the effective mass of a single 2-atom and $g$ is the strength of the effective interaction between different atoms. The prime on the sum indicates that the $\qq=0$ term is omitted, since this is included in the $\epsilon_2$ term. We shall assume that the momentum scale for variations of the coupling strength are large compared with the Fermi momentum, and therefore the momentum dependence of the coupling may be neglected.  Also, we have not written explicitly the direct interaction between two 2-atoms or between two 1-atoms because this is short-ranged and, consequently,  the direct and exchange contributions to the interaction energy cancel.  We shall consider only the isotropic part of the interaction and neglect the momentum dependence of the interaction which leads to contributions to Landau parameters other then that for $l=0$.

The coupling constant $g$ gives the change in the interaction energy between small long-wavelength density disturbances in the medium.  In the limit of small concentrations of 2-atoms the contribution to the energy from the Fermi motion of the 2-atoms  is negligible and therefore the interaction energy may be replaced by the total energy.  Thus one sees that \cite{footnotederiv}
\be
g=\frac{\partial^2{\cal E}(n_1,n_2)}{\partial n_1\partial n_2}=\frac{\partial{\epsilon_2}}{\partial n_1}=\frac{\partial{\mu_2}}{\partial n_1},
\ee
where ${\cal E}(n_1,n_2)$ is the energy density as a function of the densities of the two components.

The induced interaction is always at least of second order in $g$ and we evaluate it by calculating the  $g^2 $ contribution to the energy, which is given by
\begin{widetext}
\be
E^{(2)}=-\frac{g^2}{V^2}{\sum_{\pp_1\pp_2\qq}}' \frac{(1-f_{\pp_1+\qq})(1-f_{\pp_2-\qq})f_{\pp_2}f_{\pp_1}}{(\pp_1+\qq)^2/2m_1-p_1^2/2m_1+(\pp_2-\qq)^2/2m_2^*-p_2^2/2m_2^*},
\label{E2}
\ee
\end{widetext}
where $f_\pp$ is the particle distribution function, the index on the momentum variable indicating whether it refers to 1-atoms or 2-atoms.  
From Eq.\ (\ref{E2}) one may calculate the corresponding contribution to the Landau effective interaction between two 2-quasiparticles,
\be
f=\frac{\delta^2 (E^{(2)}/V)}{\delta f_{\pp_2}\delta f_{\pp'_2}},
\label{f}
\ee
where $\bf p_2$ and $\bf p_2'$ are taken to be vanishingly small.  Thus one finds
\be
f=\frac{g^2}{V}  \left(\sum_{\pp_1}\frac{f_{\pp_1}-f_{\pp_1+\qq}}{(\pp_1+\qq)^2/2m_1-p_1^2/2m_1}\right)_{q\rightarrow 0}=g^2{\cal N}_1(0).
\label{fFF}
\ee
This is positive, since although the contribution to the total energy is negative, one of the distribution functions corresponds to a hole line, which carries a factor $1-f_{\pp_2-\qq}$, and therefore the second functional derivative of the energy with respect to the distribution function for 2-atoms is positive.  Expressed in the language of Ref.~\cite{mora}, Pauli blocking reduces the magnitude of the negative contribution to the energy, thereby giving a  positive contribution to $f$. An equivalent description is that, while the induced interaction is intrinsically negative, its contribution to the Landau quasiparticle interaction comes from an exchange term, which gives an additional minus sign \cite{YZP}.  Contributions of higher order in $g$ will contain additional powers of the Fermi momentum of the minority component and are therefore negligible in the limit of a low concentration of minority atoms.

\section{A Bose gas with a dilute Fermi component}
An analysis similar to that in Sec.\ {\ref{Sec:FF}} may be carried through for a majority Bose component. The elementary excitations in the Bose system are phonons and the low-energy effective Hamiltonian is
\begin{align}
H=&E_0(N_1)+\sum_{\qq}  sq \alpha^\dagger_\qq   \alpha_\qq +\epsilon_2N_2 +\sum_{\pp_2}  \frac{p_2^2}{2 m_2^*} b^\dagger_{\pp_2}   b_{\pp_2}\nonumber\\
&+\frac{g}{V}{\sum_{\qq\pp_2}}' M_\qq   (\alpha^\dagger_\qq + \alpha_{-\qq})b^\dagger_{\pp_2-\qq} b_{\pp_2} ,
\label{hamBose}
\end{align}
where the operator $\alpha^\dagger_\qq$ creates phonons in the Bose system and
$E_0(N_1)$ is the energy of $N_1$ bosons  including the effects of interactions.  Again we neglect the momentum-dependent part of the effective interactions due to, e.g.,  coupling of fermions to the superfluid velocity of the bosons. The energy of a low-wavelength phonon of wavenumber $q$ in the Bose system is $sq$, where $s^2=(n_1/m_1)\partial \mu_1/\partial n_1$.  The matrix element $M_q$ for the density operator to create or destroy a phonon of momentum $q$ is given by \cite{nozierespines}
\be
M_q=\left(\frac{N_1q}{2m_1s}\right)^{1/2}.
\ee
 There are no terms with higher powers of $\alpha$ and $\alpha^\dagger$ since phonons in a Bose-Einstein condensed gas exhaust the frequency-weighted sum rule for the density--density correlation function.   The contribution to the energy calculated in second-order perturbation theory is
 \begin{align}
 E^{(2)}=&-\frac{g^2}{V^2}\sum_{\qq\pp_2} |M_q|^2 \left(\frac{(1+n_\qq)(1-f_{\pp_2-\qq})f_{\pp_2}}{(\pp_2-\qq)^2/2m_2^* +sq-p_2^2/2m_2^*}  \right.\nonumber\\
 & \left. +   \frac{n_{-\qq}(1-f_{\pp_2-\qq})f_{\pp_2}}{(\pp_2-\qq)^2/2m_2^*-sq-p_2^2/2m_2^*}      \right),
\label{E2Bose}
 \end{align}
 where $n_\qq$ is the phonon distribution function.
 Thus for two fermions on the Fermi surface, the effective interaction given by Eq.\ (\ref{f}) is
 \be
 f=\frac{g^2n_1}{2m_1 s^2}=g^2\frac{\partial n_1}{\partial \mu_1}.
 \label{fBose}
 \ee
This result is valid regardless of the strength of the boson-boson interaction. 
The only difference compared with the case of a majority gas of fermions is that ${\cal N}_1=\partial n_1/\partial \mu_1$ in Eq.\ (\ref{fFF}) must be replaced by the expression for a Bose gas.
For a weakly-interacting Bose gas, $\mu_1=n_1 U_{11}$, where $U_{11}=4\pi a_{11}/m_1$ is the effective low-energy interaction, $a_{11}$ being the scattering length, and therefore $\partial \mu_1/\partial n_1=U_{11}$.

\section{Minority Bose component}

Analogous arguments may be carried through for a mixture with a minority Bose component.  We consider the case when the bosons are in a Bose--Einstein condensate and do not treat the case where bosons form fermionic diatomic molecules with the majority fermions \cite{zhou, zhai}.  The bosons may thus be described by their density, $n_2$, and the boson superfluid velocity, ${\bf v}_s$.  For a spatially uniform  system consisting of bosons, its ground state energy is a function only of the boson density, $n_2$, so the effective interaction is defined by the relation
\be
f=\frac1V\frac{\delta^2 E^{(2)}}{\delta n_2^2}.
\label{fBose}
\ee
For moving condensates, there will also be effective interactions involving the superfluid velocity, but we shall not take these into account explicitly here.  One difference compared with the case of minority fermions is that the direct boson-boson interaction does not vanish. As defined in Eq.\ (\ref{fBose}), there is no exchange term since the 2-atoms are in a Bose condensate.  Consequently the exchange process is identical to the process without exchange and to include it explicitly would be double counting.    A second difference is that
the induced interaction contribution to $f$ is negative, because the wave function for bosons is symmetric under interchange of particles and, consequently, the extra minus sign acquired in the case of fermions is absent.
The result is
\be
f=f^{\rm dir}-g^2 \frac{\partial n_1}{\partial \mu_1}.
\label{inducedBose}
\ee

The collective modes of the Bose system may be calculated by the Bogoliubov approach in which the effective interaction between bosons is given by $f$, provided the frequency dependence of the interaction may be neglected.  This condition is satisfied provided the velocity of the Bogoliubov mode is much less than the Fermi velocity or the sound speed of the majority component.  Thus the sound speed, $s_2$, in the Bose gas is given by
\be
s_2^2=\frac{n_2 f}{m_2^*}.
\label{soundspeedBose}
\ee
If $f$ is negative, the sound speed is imaginary and density modes in the system are unstable.  This corresponds to the thermodynamic stability condition, see, e.g.,  Ref. \cite[Sec. 12.1.1]{pethick}
\be
\det \left(\frac{\partial \mu_i}{\partial n_j}\right)\ge 0,
\ee
since $f^{\rm dir}=\partial \mu_2/\partial n_2$.

\begin{figure}
  \includegraphics[width=3.2in]{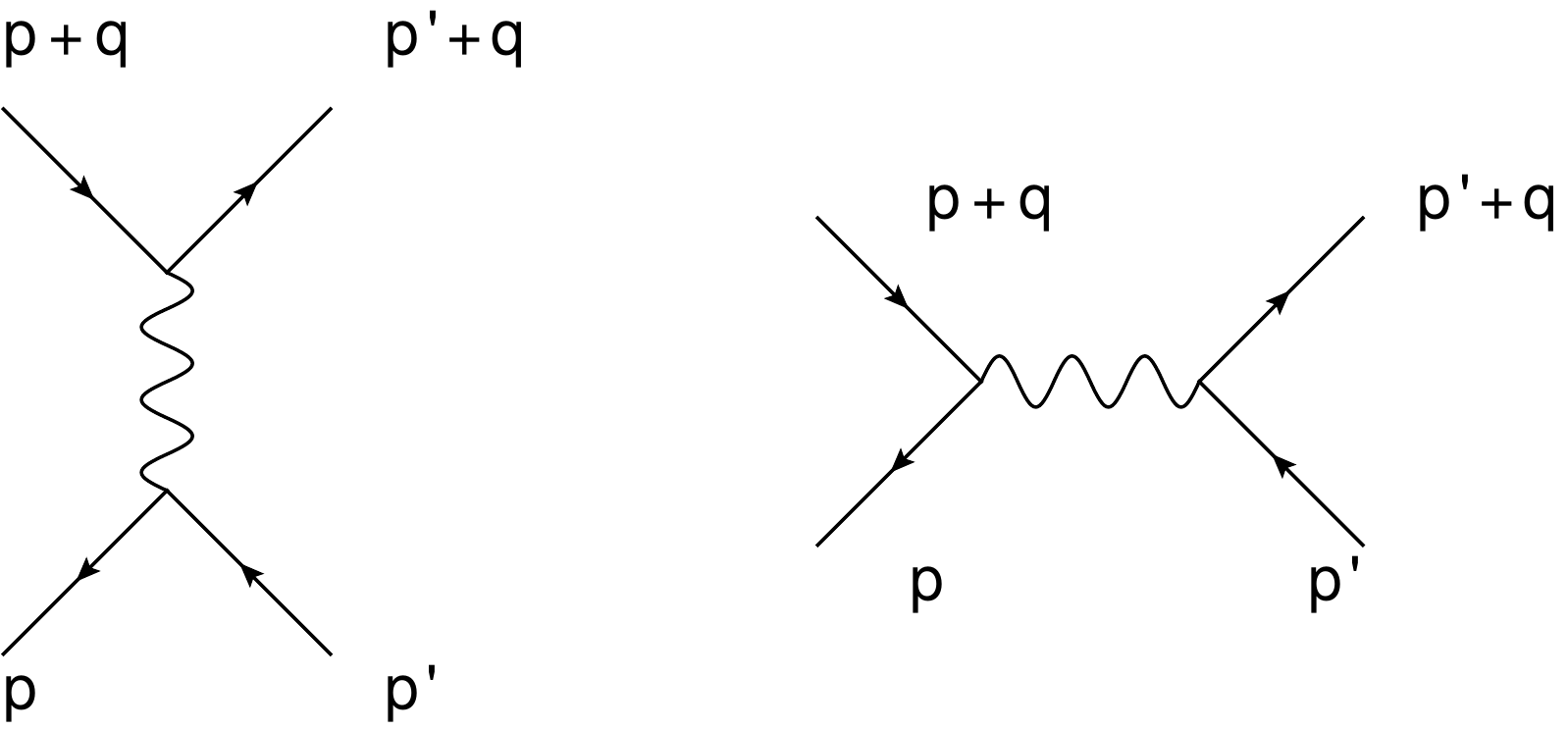}
   \caption{Diagrams representing the induced contribution to the effective interaction $f$ (left) and the corresponding diagram for the crossed  channel (right).  The wavy line represents exchange of an excitation in the majority component, either a phonon in the case of a Bose gas or a particle-hole pair in a Fermi gas.}
   \label{Induced1}
\end{figure}

\section{Dynamical effects}

Here we show how density modes are affected by the induced interaction.  For definiteness, let us consider the response of the density of a minority fermion component to a potential acting on the minority atoms.  We assume the  wave vector $\qq$ of the perturbation to be small.   The contributions to the effective interaction between minority atoms have the forms shown in Fig. (\ref{Induced1}): the left-hand diagram corresponds to the contribution to the effective interaction that we have calculated earlier, while the second term is due to the response of the majority atoms at wave vector $\qq$.  Summing up all bubble diagrams, the density--density response function for the minority atoms is given by 
\be
\chi_2(\qq,\omega)=\frac{\chi_2^{(0)}(\qq,\omega)}{\left(1+[f-f(\qq, \omega)]\chi_2^{(0)}(\qq,\omega)\right)},
\ee
where
\be
\chi_2^{(0)}(\qq, \omega)=\frac{m_2^*p_{F2}}{2\pi^2}\left(1-\frac{\omega}{2qv_{F2}}\ln\left[\frac{\omega+qv_{F2}}{\omega-qv_{F2}}\right]\right)
\label{chi1}
\ee
and
\be
f(\qq, \omega) =g^2\chi_1(\qq,\omega).
\ee
If the majority atoms are fermions, $\chi_1(\qq,\omega)$ is given by
\be
\chi_1(\qq, \omega)=\frac{m_1p_{F1}}{2\pi^2}\left(1-\frac{\omega}{2qv_{F1}}\ln\left[\frac{\omega+qv_{F1}}{\omega-qv_{F1}}\right]\right),
\label{chi1}
\ee
while if they are bosons it is given by
\be
\chi_1(\qq, \omega)=\frac{n_1q^2}{m_1}\frac1{s_1^2q^2-\omega^2}.
\label{chi1Bose}
\ee
In the limit $\omega/q \rightarrow 0$, $f(\qq, \omega)$ tends to $f$.  The cancellation of the two terms is a consequence of the antisymmetry under interchange of two fermions in the same internal state and the fact that for $\omega/q\rightarrow 0$ the contributions of the two  processes shown in Fig.\ \ref{Induced1} are equal in magnitude. Consequently, effects of the 1-2 interaction disappear in the density response.  However, at nonzero frequency, there will in general be effects due to the interaction because the ratio of the energy transfer to the momentum transfer is different in the two particle--hole channels for the 2-atoms. In words, while the Pauli principle forbids two atoms in the same internal state being at the same point in space at a given time, it does  not forbid two such particles being at the same point at different times.

The frequency dependence of the interaction has striking implications for the scattering rate of 2-atoms by 2-atoms, since for $\omega=0$ the effective interaction vanishes and consequently the scattering rate will grow with temperature as $T^4$, in contrast to the $T^2$ behavior predicted by standard Landau theory \cite{pwavefootnote}.  Related effects occur for quark--gluon plasmas, where again the frequency dependence of the scattering amplitude plays a decisive role in determining scattering rates at low temperature \cite{Heiselberg}.  However, the rate for scattering of 2-atoms by majority fermonic atoms will still have the usual $T^2$ dependence. 

For a minority Bose gas, the frequency dependence of the effective interaction affects the dispersion relation of the Bogoliubov mode, which is given by
\be
\omega^2=\frac{n_2}{m_2^*}\left[f^{\rm dir}-g^2\chi_1(\qq,\omega)\right]q^2.
\ee
For a majority Bose gas, $\chi(\qq, \omega)$ increases for small $\omega$ and consequently the sound speed is less than the result (\ref{soundspeedBose}) predicted on the basis of the static interaction (\ref{inducedBose}).  The situation for  a majority Fermi gas is different, and for small $\omega/v_1q$,
\be
\chi_1(\qq,\omega) \simeq {\cal N}_1(0)\left(1+\frac{\rmi \pi}{2}\frac{\omega}{v_{F1}q}\right).
\ee
The leading effect of dynamics at low frequencies is an imaginary contribution to $\chi_1$ due to Landau damping. Thus sound waves in the Bose gas can decay  into particle-hole pairs of the majority species.  The time $\tau$ for decay of the intensity of the wave is given by 
\be
\frac1{\tau}=-2\,{\rm  Im}\,\omega=\frac{q}{4\pi}\frac{n_2g^2 m_1^2}{m_2^*},
\ee
which is small compared with the real part of the frequency for small concentrations of the minority component.
 In Appendix A we present on the basis of a functional integral approach an alternative derivation of a number of results in this section for the case of dilute gases.

\section{Implications for dilute solutions of helium isotopes}

Our considerations above have implications for the theory of dilute solutions of $^3$He in liquid $^4$He.   The standard approach adopted by Bardeen, Baym and Pines \cite{BBP}(BBP) is to assume that the interaction between two $^3$He impurities may be modeled by a potential that is local in time but of nonzero range in space \cite{BBP}.  This potential includes effects of the induced interaction between $^3$He atoms due to exchange of excitations in the $^4$He.  The parameters of the potential are typically obtained by assuming a particular form for the spatial Fourier transform of the interaction and then fitting parameters to obtain agreement with measured transport coefficients.   In this approach, the dynamics of the $^4$He atoms is taken into account implicitly, since it is assumed that the $^4$He atoms respond on a time scale short compared with characteristic times for the $^3$He. In the Landau theory of the dilute solutions, the response of the $^4$He is taken into account through its effect on  contributions to the Landau parameters.  This effective interaction corresponds to the definition
\be
f^{\rm BBP}=\frac1V \left.\frac{\delta^2 E}{\delta f_{\pp}\delta f_{\pp'}}\right|_{\mu_4},
\ee
where $f_\pp$ is the distribution function for $^3$He quasiparticles and $\mu_4$ is the $^4$He chemical potential.  With this definition, the local density of $^4$He atoms adjusts to the local density of $^3$He atoms, and it is this interaction that should be used in formulating a theory of the static properties of mixtures, such as the magnetic susceptibility  or the $^3$He contribution to the bulk modulus, if one wishes to avoid treating explicitly the response of the $^4$He.

For calculating dynamical effects, the $^4$He density must to be treated as a dynamical variable, and therefore the appropriate definition of an effective interaction is
\be
f=\frac1V\left.\frac{\delta^2 E(f_\pp, n_4,\vv_4)}{\delta f_{\pp}\delta f_{\pp'}}\right|_{n_4,\vv_4}.
\ee
This corresponds to the natural generalization of the original Landau definition of a quasiparticle interaction to allow for a second component.  For the case of a majority Fermi gas rather than $^4$He, the density $n_4$ and the superfluid velocity $\vv_4$ would be replaced by  the distribution function for the majority component.  Dynamical processes may then be described in terms of, e.g., the kinetic equation for $^3$He quasiparticles and the equations of superfluid hydrodynamics for the $^4$He.

We now consider the relationship of the two different definitions of effective interaction to microscopic theory \cite{Saam}.
The standard definition of the quasiparticle interaction in a normal Fermi system is \cite{Nozieres}
\be
f_{\pp,\pp'}=\lim_{q\rightarrow 0} \lim_{\omega/v_F q\rightarrow \infty} z^2\Gamma(\pp,\pp',\qq,\omega),
\label{standard_f}
\ee
where $z$ is the wave function renormalization parameter, and $\Gamma(\pp,\pp',\qq,\omega)$  is the two-particle vertex function for bare particles.  The momenta   $\pp$ and $\pp'$ are those of the incoming particles, $v_F$ is the Fermi velocity, $\qq$ is the momentum transfer and $\omega$ the energy transfer.  In the case of mixtures, the vertex function depends not only on the distribution of fermions but also has contributions from interactions of fermions with bosons.  If one applies the definition (\ref{standard_f}) to helium mixtures, contributions to the vertex function for scattering of two $^3$He atoms due to exchange of phonons in the Bose system carrying momentum $\qq$ will vanish, because the phonon propagator behaves as $n_4q^2/m_4\omega^2$ (cf. Eq.\ (\ref{chi1Bose})).  In physical terms, the $^4$He does not respond because of the high frequency. For a low concentration of $^3$He,  $v_F$ is very much less than $s$.  The effective interaction used by BBP includes the effects of phonon exchange in the particle--hole channel with momentum $\qq$ and in the microscopic theory it corresponds to the quantity
\be
f_{\pp\uparrow,\pp'\downarrow}^{\rm BBP}=\lim_{q\rightarrow 0} \lim_{v_F\ll \omega/q\ll s} z^2\Gamma(\pp\uparrow,\pp'\downarrow,\qq,\omega).
\ee
In physical terms, $\omega/q$ must be much less than $s$ in order that the response of the  $^4$He atoms be given by its zero-frequency value.
 In the BBP approach, exchange is taken into account explicitly, and therefore, in defining an effective interaction, the two $^3$He atoms are taken to be in different spin states.

In the standard definition of effective interactions for a Fermi liquid, the two particle-hole channels are treated differently in the case of two fermions in the same spin state. For the channel with momentum transfer $\qq$, the  limit taken is the high frequency one, while for the channel with momentum transfer $\pp+\qq-\pp'$, it is the low frequency one, since the energy transfer is zero.  The effective interaction vanishes for $\pp=\pp'$  in the Born approximation but not when many-body  processes involving particle-hole pairs are taken into account.  This problem has been investigated in detail in Ref.\  \cite{Mermin}.

\section{Concluding remarks}  In this paper we have shown how the contribution to the effective interactions between minority atoms induced by interactions with a majority component may be expressed in terms of thermodynamic quantities.  An important remark is that the results are not limited to cases where the majority gas is weakly interacting.  Experimentally, an interesting case to explore would be a Bose gas with weak, repulsive interactions.  In this case the sound speed is small and therefore the effects of the induced interaction can be correspondingly large.

In addition to the interaction mediated by coupling to density fluctuations in the majority component, there is also an interaction between minority atoms due to coupling via a current--current coupling \cite{BBP} or, in the case of a fermion majority component, distortions of the majority Fermi surface that are not spherical.  In general these too should be included, but they are generally much smaller than the interaction induced by the density--density coupling.  This interaction does not contribute to the effective interaction between two fermions at the Fermi surface in a majority Bose component because the current operator for transitions of a fermion between two states on the Fermi surface is transverse, and therefore it cannot create a phonon in the Bose gas, which is longitudinal.\\

 We are grateful to Sascha Z\"ollner for helpful discussions in the earlier stages of this work. Part of this work was performed while we enjoyed the hospitality of the Institute for Nuclear Theory,  University of Washington, Seattle. ZY acknowledges support from the Tsinghua University Initiative Scientific Research Program, NSFC under Grant No.~11104157, NSF Grant DMR-0907366, and by DARPA under the Army Research
Office Grant Nos.~W911NF0710464 and W911NF0710576.

 \appendix
 \section{Functional integral approach to Fermi-Bose mixtures}
Here we give a derivation of some of the results  in Sec.~V based on a functional integral approach.
We consider a binary mixture consisting of a fermion species and a boson species at zero temperature and we shall assume that the densities of the two components are so low that binary interactions dominate. The Hamiltonian is 
\begin{align}
H=&\int\left\{\phi^{\dagger}\left(-\frac{\nabla^2}{2m_B}-\mu_B\right)\phi+\psi^{\dagger}\left(-\frac{\nabla^2}{2m_F}-\mu_F\right)\psi\right\}\nonumber\\
&+\frac{g_{BB}}2\int\phi^\dagger\phi^\dagger\phi\phi+g_{BF}\int\psi^\dagger\phi^\dagger\phi\psi,
\end{align}
where  the field operators are $\phi$ for the bosons and $\psi$ for the fermions and the integral is over coordinate space. The interactions between bosons are repulsive. The couplings are given by $g_{BB}=4\pi a_{BB}/m_B>0$ and \cite{zhou} 
\begin{align}
1/(m_r g_{BF})+\Lambda/{\pi^2}=1/{2\pi a_{BF}}.
\end{align}
Here the reduced mass is $1/m_r=1/m_B+1/m_F$, $\Lambda$ is the momentum cutoff, and $a_{BB}$ and $a_{BF}$ are the s-wave scattering lengths.

To simplify the discussion,  we assume that the Bose gas is dilute, in the sense  that $a_{BB}^3n_B\ll1$. In the case that the fermions are the minority and the Fermi momentum $k_F\to0$, the induced interactions between the minority fermions mediated by the majority bosons gives rise to a nonzero Landau parameter as discussed in Sec.~III.
In the case that the bosons are the minority, we shall consider situations where bound states of bosons and fermions (fermionic diatomic molecules) need not be taken into account.  This could be due to either the boson--fermion interaction being so weak that there are no bound states or to the state under consideration being a metastable one in which there are no such molecules.  When bound states are present, the perturbation theory treatment given here needs to be extended to allow for the presence of molecules.  In the path integral representation, after the fermions been integrated out and terms up to the second order of $g_{BF}$ retained, the partition function is given by
\begin{align}
\mathcal Z=\mathcal Z^{(0)}_f\int \mathcal D\phi \;\exp({S_{\rm eff}[\phi,\mu_B,\mu_F]}).
\end{align}
Here the effective action for the bosons  is given by
\begin{align}
&S_{\rm eff}=  \rmi\int d1\left\{\phi^{\dagger}(1)\left(\rmi \frac{\partial}{\partial t_1}+\frac{\nabla_1^2}{2m_B}+\mu_B-g_{BF}n^{(0)}_F\right)\phi(1)\right.\nonumber\\
-&\left.\frac12\int d2\left[g_{BB}\delta(1-2)+V_{ind}(1-2)\right]\phi^\dagger(1)\phi(1)\phi^\dagger(2)\phi(2)\right\},
\end{align}
where the symbol $1$ stands for $\{\mathbf r_1, t_1\}$. The integral over imaginary time $t$ is from $0$ to $-\rmi \beta$, and $n^{(0)}_F=(2m_F\mu_F)^{3/2}/6\pi^2$. The induced interaction has the form
\begin{align}
V_{\rm ind}(1-2)=-\rmi g_{BF}^2 G^{(0)}_F(1,2)G^{(0)}_F(2,1),  \label{bfvind}
\end{align}
where $G^{(0)}_F$ is the free fermion Green function.

We apply the Bogoliubov approximation for the bosons and obtain for the frequency $\Omega$ of bosonic modes the dispersion relation  
\begin{align}
\Omega^2=\frac{P^2}{2m_B}\left(\frac{P^2}{2m_B}+2n_B[g_{BB}+\tilde V_{\rm ind}(P,\Omega)]\right),
\end{align}
where   $P$ is the momentum of the mode and $\tilde V_{\rm ind}$ is the Fourier transform of Eq.~(\ref{bfvind}).
In the long wavelength limit $P/k_F\to0$,
\begin{align}
\tilde V_{\rm ind}(P,\Omega)\approx-g_{BF}^2\frac{\partial n^{(0)}_F}{\partial\mu_F}\left\{1-\frac{\Omega}{2Pv_F}\log\left[\frac{\Omega/Pv_F+1}{\Omega/Pv_F-1}\right]\right\}.
\end{align}
Here $v_F=k_F/m_F$ is the Fermi velocity. The sound speed $c\equiv \Omega/P$ for $P\to0$ and $\Omega\to0+\rmi \delta$ and in the regime $c/v_F\ll1$ is
\begin{align}
{\rm Re}\,c=&c_0\left(1-\frac{g_{BF}^2}{g_{BB}}\frac{\partial n^{(0)}_F}{\partial\mu_F}\right)^{1/2}\label{rec}\\
{\rm Im}\,c=&-\frac{\pi g_{BF}^2n_B}{4v_F}\frac{\partial n^{(0)}_F}{\partial\mu_F},\label{imc}
\end{align}
with $c_0=\sqrt{g_{BB}n_B/m_B}$.
The reduction of the sound speed, ${\rm Re}\,c<c_0$, is due to the attractive induced interactions between bosons mediated by fermions \cite{yip, giorgini}.
The small imaginary part of $c$ indicates that the  modes decay into particle--hole pairs of the majority fermions (Landau damping).

Equation (\ref{rec}) implies dynamic instability for $g_{BB}-\nobreak g_{BF}^2\partial n^{(0)}_F/\partial\mu_F\le0$. The same instability condition can be deduced from the energy density of the mixture
\begin{align}
\mathcal E=\frac{6^{5/3}\pi^{4/3}n_F^{5/3}}{20m_F}+g_{BF}n_Bn_F+\frac12g_{BB} n_B^2,\label{emean}
\end{align}
where the first term is the energy of a filled Fermi sea, by requiring that the variation of $\mathcal E$ to second order in the density variations be negative \cite{pethick}. Note that Eq.~(\ref{emean}) takes into account the interaction energy only at the mean field level.

When $g_{BF}$ is no longer small, a similar dynamic instability condition can be derived following the above argument. We can still formally integrate out the fermions in the partition function and obtain an effective action for the bosons.  Since $n_B\to0$, in the effective action for the bosons, the interaction effects due to $g_{BF}$ can be taken into account in two steps. The first step is to modify the properties of a single boson, such as the effective mass $m_B^*$, the single boson energy $\mu_P$ (as for polarized fermions) and the quasiparticle residue $z(<1)$. The second step is to change the interactions between these bosonic quasiparticles. In the low energy and long wavelength limit, induced interactions between the bosons are given by the diagrams shown in Fig.~(\ref{Induced1}) with the vertex replaced by $\partial \mu_P/\partial n_F^{(0)}$. Within the Bogoliubov approximation, we conclude that if
\begin{align}
g_{BB}-\left(\frac{\partial\mu_P}{\partial n_F^{(0)}}\right)^2\left(\frac{\partial n^{(0)}_F}{\partial\mu_F}\right)\le0,\label{binstable}
\end{align}
the system becomes dynamically unstable.


\begin{thebibliography}{99}
\bibitem{footnotederiv} Unless stated otherwise, a partial derivative with respect to the density of one component is taken at fixed density of the other component, cf. Eq.\  (\ref{nu}).
\bibitem{pwavefootnote}  If one includes momentum terms in the effective interactions, there will be contributions to scattering rates that vary as $T^2$ but their magnitude will be suppressed by powers of the density of the  minority component.
\end{thebibliography}
\end{document}